\documentclass[aps,prl,twocolumn,groupedaddress,amsmath,amssymb,amsfonts]{revtex4}

\DeclareMathOperator{\tr}{tr}

\newcommand{\bracket}[1]{\langle #1 \rangle}
\newcommand{\ket}[1]{\rvert #1 \rangle}

\usepackage{graphicx}
\usepackage{bm}
\usepackage{bbold}
\usepackage{wasysym}
\newcommand{\svec}[1]{\bm{#1}}
\renewcommand{\vec}[1]{\mathbf{#1}}
\newcommand{\vac}{\lvert \text{vac} \rangle}

\begin{document}

\title{Fractal valence bond loops in a long-range Heisenberg model at criticality}

\author{K.\ S.\ D.\ Beach}
\email[]{ksdb@physik.uni-wuerzburg.de}
\affiliation{Institut f\"{u}r Theoretische Physik und Astrophysik, Universit\"{a}t W\"{u}rzburg, Am Hubland, D-97074 W\"{u}rzburg, Germany}

\date{July 2, 2007}

\pacs{05.45.Df, 75.10.Jm, 75.10.Nr, 75.30.Ds, 75.40.Mg, 75.40.Cx}

\begin{abstract}
We present a valence bond theory of the spin-$S$ quantum Heisenberg model.
For nonfrustracting, local exchange and dimension $d > 1$, it predicts a resonating ground 
state with bond amplitudes $h(r) \sim (a^2+r^2)^{-p/2}$ and decay exponent $p=d+1$. 
Different values of $p$ can be achieved by introducing frustrating ($p > d+1$) 
or nonfrustrating  ($p < d+1$) long-range interactions. For $d=2$, but \emph{not} $d=3$, 
there is a critical value of the decay exponent $p_{\text{c}}$ above which the
ground state is a spin liquid. The phase transition is analogous to quantum percolation,
with fractal valence bond loops playing the role of percolating clusters.
The critical exponents are continuously tunable
along the phase boundary $p=p_{\text{c}}(a,S)$.
\end{abstract}

\maketitle

Singlet product states were introduced in the early days of quantum chemistry 
to describe valence bonding in molecules~\cite{Rumer32,Pauling33}.
These ``valence bond states'' were also applied to translationally invariant
systems and proved especially useful for studying interacting $S=\tfrac{1}{2}$ quantum spin
chains~\cite{Hulthen38,Majumdar69}.
A featureless resonating valence bond (RVB) wavefunction, consisting of a
superposition of all possible configurations of short range valence bonds,
was later promoted by Anderson as a possible ground state for 
low-coordination-number antiferromagnets~\cite{Anderson73, Fazekas74}
and as a potential route to superconductivity in the cuprates~\cite{Anderson87a,Kivelson87a}.

Liang, Doucot, and Anderson (LDA) extended the RVB picture to include
bonds on all length scales 
by assuming that the weight associated with
each configuration could be factorized into a product of individual bond
amplitudes, each expressed as a single function of the bond length~\cite{Liang88}.
They considered various functional forms 
for the two-dimensional Heisenberg model and 
concluded that the optimal amplitudes 
fall off as a powerlaw: $h(r) \sim r^{-p}$.
Their variational calculation was not sufficiently accurate 
to resolve the correct exponent, which is 
almost certainly $p=3$~\cite{Sandvik05,Lou06,Havilio99}.

In this Letter, we begin by providing a formal justification for the LDA ansatz.
The calculation presented here is a mean field treatment of the
Heisenberg model based on valence bond creation and annihilation operators
\cite{Beach06}; these obey an unusual operator algebra that reproduces the properties
of the overcomplete valence bond basis. 
We show that a broad class of spin-$S$ models with two-spin
interactions have ground state wavefuctions that are very close to being
factorizable-amplitude RVB states. Antiferromagnetically ordered states
attain perfect factorizability in the large spin limit.

For models with local, nonfrustrating interactions, the mean field equations can be
solved analytically. The amplitudes are generically radially symmetric and of the form
$h(r) = (a^2 + r^2)^{-p/2}$, where the core size of the distribution is $a = 1/\sqrt{d}$
and the powerlaw tail has exponent $p = d+1$. Here $d$ is the dimension of the lattice.
The only exception is for $d=1$, in which case either the bond lengths are gaussian 
distributed (even $2S$) or beyond mean field prediction (odd $2S$). In summary,
\begin{alignat}{2} \label{EQ:hr1D2Seven}
h(r) &= \frac{2}{\sqrt{\pi}} \biggl[  \frac{\xi-1}{\xi(\xi+1)} \biggr]^{\!1/2} \!e^{-r^2/2\xi^2} & \quad d&=1, \, \text{$2S$ even}\\ \label{EQ:hr2D}
h(r) &= \frac{1}{\pi(\frac{1}{2} + r^2)^{3/2}} & d&=2\\ \label{EQ:hr3D}
h(r) &= \frac{2\sqrt{6}}{3\pi^2(\frac{1}{3} + r^2)^2} & d&=3
\end{alignat}
The $\xi$ appearing in Eq.~\eqref{EQ:hr1D2Seven} is the 
spin correlation length
\begin{equation}
\xi^2 = 1 + \frac{1}{\pi^2} e^{\pi(2S+1)}.
\end{equation}

We then ask the question, are there couplings $J_{ij}$ that
preserve the $h(r) = (a^2 + r^2)^{-p/2}$ form but accommodate
arbitrary values of $p$\,? Indeed, it turns out that tuning $p$ away 
from $d+1$ is equivalent to introducing a long-range, bipartite interaction
($i,j$ in opposite sublattices)
\begin{equation} \label{EQ:LongRangeInteractions}
J_{ij} = \gamma_{ij} + \bigl(1-\gamma_{ij}\bigr)\frac{\beta(p)(d+1-p)}{r_{ij}^{\alpha(p)}},
\end{equation}
where $\gamma_{ij}$ is the nearest neighbour matrix and
$\alpha(p), \beta(p)$ are smooth, positive functions.
The beyond-nearest-neighbour part 
is frustrating when $\delta p = p-d-1>0$.
For the square lattice, $\alpha(3+\delta p) = 4 + 0.84\,\delta p$
and $\beta(3+\delta p) = 0.7 - 0.15\,\delta p$.

\begin{figure}
\includegraphics{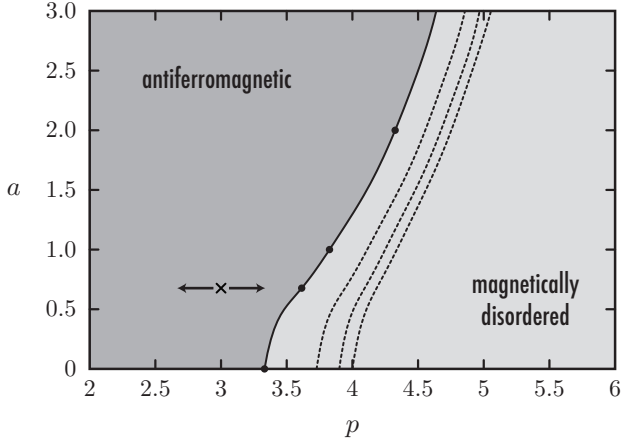}
\caption{ \label{FIG:criticalp} 
A phase diagram for the RVB wavefunction  with amplitudes $h(r) = (a^2+r^2)^{-p/2}$
on the square lattice.
$r$ is the length of a valence bond connecting spins in opposite sublattices.
The phase boundary is drawn for $S=\tfrac{1}{2}$ (solid line) and for $S=1,\tfrac{3}{2},2$ 
(dotted lines, from left to right). The location of the ground state ($\times$) 
can be adjusted horizontally by perturbing [as per Eq.~\eqref{EQ:LongRangeInteractions}]
about the nearest-neighbour Heisenberg model. Points at which the
critical exponents have been measured (see Table~\ref{TAB:exponents})
are marked ($\bullet$).
}
\end{figure}

The function $h(r) = (a^2 + r^2)^{-p/2}$ characterizes
a family of RVB states interpolating between the classical N\'{e}el state ($p\to 0$)
and Anderson's short-range RVB state ($p \to \infty$).
For $d=2$, these two limits are separated by a phase transition: there is a critical 
line $p = p_{\text{c}}(a,S)$ beyond which the system is quantum disordered.
See Fig.~\ref{FIG:criticalp}. 
Along the critical line, the valence bond loops are scale invariant and have
fractal dimension $D_f$. Their length distribution is characterized by
entropy and loop tension exponents, $\tau(a,S)$ and $\sigma(a,S)$, 
in terms of which all other critical exponents can be expressed.
In higher dimensions, where even the short-range RVB state is N\'{e}el ordered,
there is no transition.

\emph{Mean field}---The 
valence bond creation and annihilation operators $\chi^{\mu\dagger}_{ij}$ and 
$\chi^{\mu}_{ij}$, discussed in Ref.~\onlinecite{Beach06}, form an overcomplete
basis for any system consisting of an even number of SU(2) spins. 
The operator $\chi^{\mu\dagger}_{ij} = (\chi^{0\dagger}_{ij},\svec{\chi}^{\dagger}_{ij})$ 
creates a singlet ($\mu=0$) or triplet ($\mu=1,2,3$) pair at sites $i$ and $j$ out of the
spinless vaccuum. More precisely,
\begin{equation}
\chi^{\mu\dagger}_{ij} \vac = \sum_{s,s' = \uparrow,\downarrow} 
\tau^{\mu}_{s,s'} \lvert s \rangle_i \otimes \lvert s' \rangle_j,
\end{equation}
where $\tau^{\mu} = ( \mathrm{i}\sigma^2, \mathrm{i}\sigma^3, \mathbb{1}, -\mathrm{i}\sigma^1 )$ 
is a vector of $2\times 2$ matrices related to the Pauli matrices by $\tau^0 = \mathrm{i}\sigma^2$ 
and $\svec{\tau} = \svec{\sigma}\sigma^2$.

Valence bond operators with no indices in common commute. Those with 
matching site indices obey a bosonic commutation relation, whereas
those with only one index in common obey a more complicated rule:
\begin{align}
\label{EQ:commutationrule1}
\bigl[ \chi^{\mu}_{ij}, \chi^{\nu \dagger}_{ij} \bigr] \vac &= \delta^{\mu\nu}\vac, \\ 
\label{EQ:commutationrule2}
\bigl[ \chi^{\mu}_{ij}, \chi^{\rho \dagger}_{kj}
\chi^{\theta \dagger}_{il} \bigr] \vac 
&= \frac{1}{2} \sum_{\nu} T^{\nu\rho\mu\theta} \chi^{\nu \dagger}_{kl}\vac.
\end{align}
Here, $T^{\nu \rho \mu \theta} = \frac{1}{2} \tr \tau^{\nu\dagger} 
\tau^\rho \tau^{\mu\dagger} \tau^\theta$ takes the values $0,\pm 1$. 
Equation~\eqref{EQ:commutationrule2} encodes the overcompleteness
of the basis.

Valence bond operators can be used to construct higher-spin states,
so long as there are exactly $2S$ bonds emerging from
each site: $\sum_\mu \chi^{\mu\dagger}_{ij} \chi^{\mu}_{ij} = 2S$.
Then each valence bond state is characterized by a set $v$ 
specifying the endpoints and singlet/triplet character of each bond. That is,
$\lvert v \rangle = \mathcal{S}\bigl[ \prod \chi^{\mu\dagger}_{ij}\bigr] \vac$,
where the product is over all $(i,j,\mu) \in v$ and $\mathcal{S}$ denotes
a symmetrization of the operators with respect to all orderings~\cite{Affleck87,Liang90}.

Since singlet bonds have a direction associated with them ($\chi^{0}_{ji} = -\chi^{0}_{ij}$, whereas
$\svec{\chi}_{ji} = \svec{\chi}_{ij}$), we restrict our discussion to bipartite lattices
($N$ spins in each sublattice) and adopt the convention that all bonds
originate in the $A$ sublattice and terminate in the $B$ sublattice.
Hence, operator indices appear in the standard order $\chi^{\mu\dagger}_{ij}$ 
with $i\in A$ and $j \in B$, and 
the transform
$\chi^{\mu}_{\vec{q}}$ 
is defined at only $N$ wavevectors in a reduced Brioullin zone.

An arbitrary state can be expressed as
a linear combination (not unique) $\lvert \psi \rangle = \sum_v \psi(v) \ket{v}$.
Factorizability in the LDA sense means that the coefficient 
$\psi(v) = \prod a_{ij}^\mu$
can be decomposed into a product of individual bond amplitudes $a^{\mu}_{ij}$.
Even under the assumption of perfect factorization, this wavefunction is nontrivial,
since there are geometrical constraints implicit in tiling the lattice
with hard-core dimers.

Those constraints can be made explicit in the following way. The bond
configuration sum can be achieved by exponentiation
of a weighted one-bond operator:
\begin{equation} \label{EQ:psiGh}
\lvert \psi \rangle 
= \hat{G} \exp \sum_{i \in A} \sum_{j \in B} \sum_\mu a^{\mu}_{ij} \chi^{\mu\dagger}_{ij}\vac.
\end{equation}
The Gutzwiller projection operator $\hat{G} = \prod_{i\in A} \delta( \hat{N}_i - 2S )$ filters out all configurations that do not have exactly $2S$ bonds emerging from each lattice site. 
This is similar in spirit to the construction of Anderson's projected BCS 
wavefunction~\cite{Anderson87a}, but here the underlying operators are not paired fermions.
The projection step can be implemented by introducing a gauge field at each
bond endpoint: $\chi^{\mu\dagger}_{ij} \to e^{\mathrm{i}\pi(\phi_i+\phi_j)} \chi^{\mu\dagger}_{ij}$
and $\hat{G} = \int\!D[\phi]\,e^{-\mathrm{i}\sum_j 2\pi S\phi_j }$.

Equation~\eqref{EQ:psiGh} can be written compactly as 
$\ket{\psi} = \hat{G}\ket{a}$ with
$\ket{a} = \exp\sum_{\vec{q},\mu} a_{\vec{q}}^\mu \chi^{\mu\dagger}_{\vec{q}} \vac$. 
If we relax $\hat{G}$ and enforce the number constraint on average
(fixing the overall bond number at $2SN$),
then $\ket{a}$ can be viewed as a mean field approximation to $\ket{\psi}$.
Since $\hat{N}_i \sim \partial/\partial \phi_i$, the total 
bond number operator $\hat{N} = \sum_{i\in A} \hat{N}_i$
is related to the overall normalization of the bond amplitudes:
\begin{equation} \label{EQ:BondNumberOperator}
\hat{N} \lvert a \rangle = \frac{\partial}{\partial u} \lvert ua\rangle \Bigr\rvert_{u=1}
= \sum_{\vec{q},\mu} a^{\mu}_{\vec{q}}\chi^{\mu\dagger}_{\vec{q}} \lvert a \rangle.
\end{equation}

Let us assume that the spin-spin interactions in the Hamiltonian
act only between sites in opposite sublattices:
$\hat{H} = \sum_{i \in A}^{j \in B} J_{ij} \vec{S}_i\cdot\vec{S}_j - \lambda(\hat{N}-N)$.
We have introduced a Lagrange multiplier $\lambda$ coupled to the shift in bond number.
In the state $\ket{a}$, the valence bond operator has an anomalous
expectation value $\Delta^\mu_{ij} = J_{ij} \bracket{\chi^\mu_{ij}}$.
For a singlet ground state, the various triplet components vanish:
$\Delta^\mu_{\vec{q}} = \Delta^0_{\vec{q}}\delta^{\mu,0} \sim JS\delta^{\mu,0}$
and $a^{\mu}_{\vec{q}} = h_{\vec{q}} \delta^{\mu,0}$.
The expectation value of $\vec{S}_i\cdot\vec{S}_j$ between valence bond states
depends on whether $i$ and $j$ are in the same valence bond loop~\cite{Liang88}.
At the mean field level, 
\begin{equation} \label{EQ:SidotSjX}
\mathcal{C}_{ij} = (-1)^{i+j}\bigl\langle \vec{S}_i\cdot\vec{S}_j\bigr\rangle 
= \frac{1}{2}\bigl\lvert X_{ij} \bigr\rvert^2 -(S+1)\delta_{ij},
\end{equation}
where 
$X_{ij}=\mathsf{F}[ h_{\vec{q}}/\bigl(1-\tfrac{1}{2}\lvert h_{\vec{q}}\rvert^2\bigr)]$
and
$X_{ik} = \mathsf{F}[\sqrt{2}/\bigl(1-\tfrac{1}{2}\lvert h_{\vec{q}}\rvert^2\bigr)]$
(with $i,k \in A$ and $j\in B$)
measure the probability that two sites are connected by a chain of bonds.
By the variational principle, $\partial\bracket{H}/\partial h_{\vec{q}}^*=0$ yields
\begin{equation}
\langle \chi^{0}_{\vec{q}} \rangle = \frac{h_{\vec{q}}}{1 - \frac{1}{2}\lvert h_{\vec{q}} \rvert^2}
= \frac{\Delta_{\vec{q}}^0}{\omega_{\vec{q}}},\ \
\omega_{\vec{q}} = \sqrt{\lambda^2 - 2|\Delta^0_{\vec{q}}|^2},
\end{equation}
and $\partial\bracket{H}/\partial \lambda =0$ yields
\begin{equation} \label{EQ:BondNumberConstraint}
\frac{1}{N}\sum_{\vec{q}} \frac{\lambda}{\omega_{\vec{q}}} = 1+2S.
\end{equation}

If $J_{ij} = J\gamma_{ij}$, the dispersion is 
$\omega_{\vec{q}} = (\epsilon^2 + 1 -\gamma_{\vec{q}}^2)^{1/2}$,
in units where $|\Delta_{\vec{q}=0}^0|=1/\sqrt{2}$ and
$\lambda^2 = 1 + \epsilon^2$.
In $d=1$, there is an important difference 
between integer and half-integer values of $S$~\cite{Haldane85}.
For even $2S$, excitations are gapped and Eq.~\eqref{EQ:BondNumberConstraint}
reads $-\frac{2}{\pi}\log \frac{\epsilon}{\pi} = 2S+1$, which leads to Eq.~\eqref{EQ:hr1D2Seven}.
For odd $2S$, the mean field breaks down; the system is gapless~\cite{Lieb61},
but $\omega_q \sim q$ causes the bond constraint equation to diverge.
[The bosonization result, $X_{ij} \sim \sqrt{\mathcal{C}_{ij}} \sim r_{ij}^{1/2}(\log r_{ij})^{1/4}$
suggests $h(r) \sim r^{-3/2}(\log\log r + O(1))$].
For $d\ge 2$, however, the gap $\epsilon = O(L^{-d})$ vanishes faster
than the wavevector spacing $\pi/L$. Thus, as $L \to \infty$, 
the divergence in the bond number equation is confined to the zero mode 
and the stucture of the equation mimics that of 
so-called ``sublattice-symmetric spinwave theory''~\cite{Hirsch88}:
\begin{equation}
\frac{1}{N}\sum_{\vec{q}} \frac{\lambda}{\omega_{\vec{q}}} \to  \frac{1}{N\epsilon} + c = 1+2S.
\end{equation}
The details of the lattice enter as a single geometrical constant,
$c \doteq 1.393$ (square), $c \doteq 1.1567$ (cubic), 
etc.
The staggered moment extracted from the large separation limit of
$X_{ij} = \bracket{\chi^0_{ij}}$ is $M = S - \frac{1}{2}(c-1)$,
the usual spinwave result~\cite{Anderson52}.
$\epsilon\to 0$ leads to the long-wavelength behaviour
$\log(h_{\vec{q}}/\sqrt{2}) = -q/\sqrt{d} + O(q^3)$, with
deviations from radial symmetry not appearing until $O(q^4)$.
The corresponding real-space bond amplitudes for $d=2,3$
[Eqs.~\eqref{EQ:hr2D} and \eqref{EQ:hr3D}] are found by
computing the Hankel transform.
In general, the presence of gapless, linearly dispersive excitations
implies $h(r) \sim 1/r^{d+1}$.
The core size $a=1/\sqrt{d} = \sqrt{2}v_{\text{s}}/\lvert \Delta^0_{\vec{q}=0}\rvert$ is
proportional to the spin-wave velocity $v_{\text{s}}$.

\begin{figure}
\includegraphics{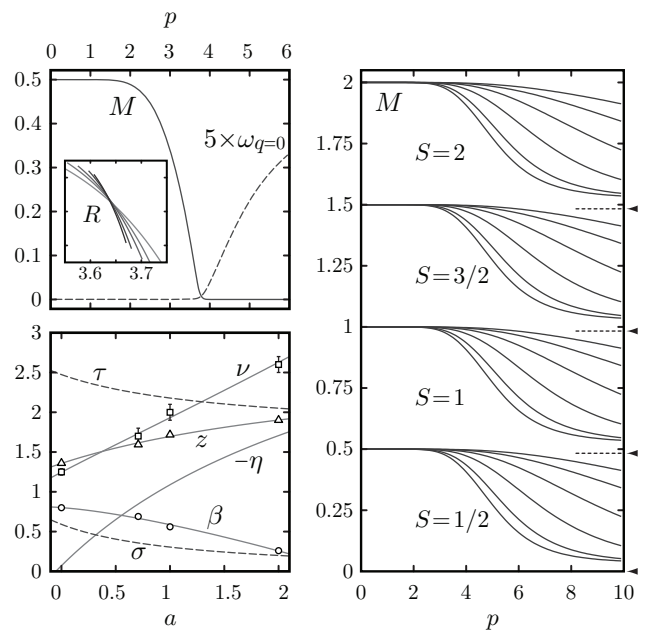}
\caption{ \label{FIG:mag} (Top-left) The staggered moment $M$
and spin gap $\omega_{\vec{q}=0}$ on a square lattice of size $L=192$
are plotted for $2S=1$, $a=1/\sqrt{2}$. 
The critical value $p_{\text{c}} = 3.6256(5)$ is determined from crossing
points of the ratio $R = X(\tfrac{L}{2},\tfrac{L}{2})/X(\tfrac{L}{4},\tfrac{L}{4})$;
the curves $L=64,96,128,192,256$ are shown in the inset.
(Bottom-left) Critical exponents $\nu$ ($\square$), $\beta$ ($\fullmoon$), and $z$ ($\triangle$) 
for $2S=1$ are plotted alongside the predicted values (solid lines) of the exponents
based on a [1,1]-Pad\'{e} approximant fit of $\tau$ and $\sigma$ (dashed lines).
(Right) The staggered moment on the $L=32$ cubic lattice is shown for 
$2S = 1,2,3,4$ and $2a = 0,1,\ldots, 5$. In the limit
$L,p,1/a \to \infty$, its value bottoms out at $M_{\infty} = S - 0.52$,
indicated by dotted lines and an arrow.
}
\end{figure}

We now consider the broader class of wavefunctions with 
amplitudes $h(r) = \bigl( a^2 + r^2 \bigr)^{-p/2}$
and compute the corresponding $p$--$a$ phase diagram.
In the $d=2$ case (Fig.~\ref{FIG:criticalp}), there is a quantum disordered phase at sufficiently large $p$ 
for all values of $a$ and $S$. 
The critical exponents ($\beta,\nu,z$ can be measured
directly; see Fig.~\ref{FIG:mag})
vary continuously
and monotonically along the line of transitions with
$\beta\!\nearrow\!0.8$, $\nu\!\searrow\!1.25$, and $z\!\searrow\!1.36$ as $a\!\searrow\!0$.
The loss of long range order can be attributed to a qualitative change in
the distribution of valence bond loops---a vanishing number of system-spanning loops
on the magnetic side becomes a macroscopic number of small loops 
on the spin liquid side~\cite{Beach06}.
At criticality, the problem is formally equivalent to that of percolation, with self-similar 
valence bond loops of fractal dimension $D_{f} = d-\beta/\nu$~\cite{Gefen81}
playing the role of percolating clusters (although loops, unlike clusters, are all backbone, i.e.,
free of dangling spins~\cite{Wang06}); fluctuations of the loop gas substitute for
the geometric average over disorder configurations.
More precisely, the analogy is with quantum, 
rather than classical percolation: the bare dynamical exponent $z_0 = 1$ should
be renormalized according to 
$z=\phi D_{f} = D_{f}$, where $\phi = z_0/(2-z_0) = 1$~\cite{Vojta05}.
This is borne out by measurements of the critical exponents, listed in Table~\ref{TAB:exponents}.

Near criticality, the loops are distributed according to
$n_l \sim l^{-\tau}\exp[-(p_{\text{c}}-p)^{1/\sigma}l]$, $n_l$ being the number of loops
of length $l$. The distribution is parameterized
by an entropy exponent $\tau$ and a loop tension exponent $\sigma$,
which are related to the conventional exponents by
$\nu=(\tau-1)/d\sigma$,
$\beta=(\tau-2)/\sigma$, and
$z=D_f=d/(\tau-1)$.
The bottom-left panel of Fig.~\ref{FIG:mag} illustrates the excellent
agreement between measured exponents and their predicted values based
on a [1,1]-Pad\'{e} fitting form for $\tau$ and $\sigma$.
Note that the anomalous dimension, determined from the relation $\eta = 2\beta/\nu -z+2-d$,
is strictly negative and goes as $\eta \approx -a$.

On the disordered side of the transition, the system is a gapped spin liquid.
In the $p \to \infty$ limit, the short-range RVB state on
the square lattice has a spin correlation function
\begin{equation}
\mathcal{C}_{ij} \sim S(S+1) \frac{e^{-r_{ij}/\xi}}{r_{ij}},
\end{equation}
where $\xi(\tfrac{1}{2}) \doteq 0.83$, $\xi(1) \doteq 2.0$, $\xi(\tfrac{3}{2}) \doteq 4.4$,
and $\log \xi(S) \sim 1.5\,S$ for large spin. 
In contrast, long-range antiferromagnetic order survives in $d=3$ 
for all $p$. This is true at the mean field  level provided that $S > \tfrac{1}{2}$
(see Fig.~\ref{FIG:mag});
a direct evaluation of the short-range RVB ground state 
(using an efficient valence bond worm algorithm)
confirms that the $S=\tfrac{1}{2}$ case is also N\'{e}el ordered.

\begin{table}
\caption{\label{TAB:exponents} Critical exponents --- $S=\tfrac{1}{2}$, square lattice}
\begin{ruledtabular}
\begin{tabular}{ccccc|c}
$a$ & $p_{\text{c}}$ & $\nu$ & $\beta$ & $z$ & $D_f$   \\\hline
0 & 3.3677(2) & 1.25(4) & 0.80(2) & 1.361(7) & 1.36(3)\\
$1/\sqrt{2}$ & 3.6256(5) & 1.7(1) & 0.69(4) & 1.594(6) & 1.59(3) \\
1 & 3.800(1) & 2.0(1) & 0.56(3) & 1.718(6) & 1.72(2)\\
2 & 4.24(1) & 2.6(1) & 0.26(1) & 1.902(4) & 1.90(5)
\end{tabular}
\end{ruledtabular}
\end{table}

\begin{figure}
\includegraphics{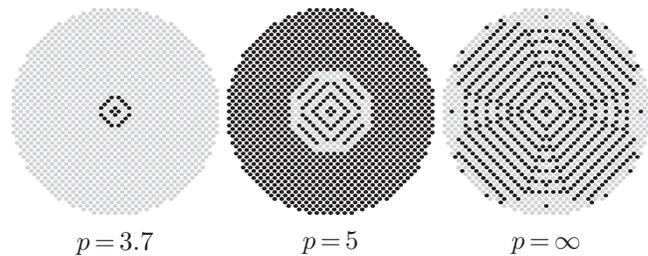}
\caption{ \label{FIG:Jpattern} The sign of the exchange interaction $J_{ij}$ is plotted for
$a = 1/\sqrt{2}$ and various values of $p>p_{\text{c}}$. Dark dots denote
$J_{ij} > 0$ and light dots $J_{ij} < 0$.
}
\end{figure}

We now run the mean field equations in reverse and solve, via
\begin{equation}
\frac{\Delta_{\vec{q}}^0}{\lambda} = \frac{h_{\vec{q}}}{1 + \frac{1}{2}{\lvert h_{\vec{q}}\rvert^2}},\quad
\frac{J_{ij}}{\lambda} 
= \frac{\mathsf{F}[\Delta_{\vec{q}}^0/\lambda]}{\mathsf{F}[\bracket{\chi^0_{\vec{q}}}]},
\end{equation}
for the interaction that stabilizes ground states with $a=1/\sqrt{d}$ and arbitrary $p$ 
(parameterizing, e.g., the horizontal trajectory in Fig.~\ref{FIG:criticalp}).
The result is the interaction given in Eq.~\eqref{EQ:LongRangeInteractions}.
Numerical estimates on the square lattice show that the long-range part 
goes as $[0.83(4-\alpha)+0.21(4-\alpha)^2+\cdots]/r_{ij}^\alpha$.
The solutions are characterized by
$\omega_{\vec{q}} \sim 1/\langle\chi^0_{\vec{q}}\rangle \sim q^\theta$
and $1-\Delta^0_{\vec{q}} \sim q^{2\theta}$,
where $\theta = 1 + (\alpha-4) + O(\alpha-4)^2 \approx \alpha-3$.
For nonfrustrating interactions ($\alpha < 4$), the dispersion is sublinear
and for frustrating interactions ($\alpha > 4$), it is superlinear.
As the excitations soften, there is a point ($d=2$ only) at which the
zero mode of the constraint equation ceases to hold macroscopic
weight. Here, the magnetic order vanishes and a gap opens in the
excitation spectrum. For $S=\tfrac{1}{2}$, that point occurs at
$\theta_{\text{c}} \doteq 1.5$ and at large $S$,
$\theta_{\text{c}} \approx 2 - \tfrac{1}{4\pi^2S}$.
Oddly, Eq.~\eqref{EQ:LongRangeInteractions} (with $J_{ij} \sim -3.6/r_{ij}^{4.6}$)
takes us up to but not through this critical point. Entry into the $p>p_{\text{c}}$ phase 
corresponds to the development of an unusual concentric-ringed, 
cloverleaf pattern of alternating frustrating and nonfrustrating interactions,
as shown in Fig.~\ref{FIG:Jpattern}. 
Other Hamiltonians with tunable interaction strengths potentially lead to
different trajectories in the $p$--$a$ plane.

A key point is that the long-range, frustrating terms in Eq.~\eqref{EQ:LongRangeInteractions}
push $p$ to higher values (toward the disordered phase) while preserving 
the radial symmetry and positivity of $h(r)$. This is generally not true of short-range frustrating 
interactions, which tend to break the radial symmetry ($h(r) \to h(\vec{r})$) and, at large 
coupling strength, the Marshall sign rule ($\exists\,\vec{r}$ such 
that $h(\vec{r}) < 0$)~\cite{Lou06}. Since reduction of the continuous rotational symmetry to the discrete 
rotational symmetry of the lattice is a precursor to crystalline
bond ordering, preservation of the radial symmetry is important
if the liquid state is not to be preempted by a bond ordered one.

The author gratefully acknowledges helpful discussions with
Anders Sandvik, Valeri Kotov, and Fakher Assaad.

\end{document}